\documentclass[prd,preprint,floatfix,nofootinbib,preprintnumbers,superscriptaddress,showkeys]{revtex4} %,twocolumn
\pdfoutput=1
\usepackage[caption=false]{subfig}
\usepackage{amsmath}
\usepackage{amsfonts}
\usepackage{amssymb}
\usepackage{float}
\usepackage{color}
\usepackage{graphicx}
\usepackage{graphics}
\usepackage{hyperref}
\usepackage{adjustbox,array}
\usepackage{enumitem} 
\hypersetup{}
\usepackage[utf8x]{inputenc} 
\usepackage{multirow}
\usepackage{booktabs}
\usepackage{mathrsfs}
\usepackage{diagbox}
\usepackage[english]{babel}
\usepackage[autostyle]{csquotes}

\graphicspath{{./figures/}}

\definecolor{rosso}{cmyk}{0,1,1,0.4}
\definecolor{rossos}{cmyk}{0,1,1,0.55}
\definecolor{rossoc}{cmyk}{0,1,1,0.2}
\definecolor{blu}{cmyk}{1,1,0,0.3}
\definecolor{blus}{cmyk}{1,1,0,0.6}
\definecolor{bluc}{cmyk}{1,1,0,0.1}
\definecolor{verde}{cmyk}{0.92,0,0.59,0.25}
\definecolor{verdec}{cmyk}{0.92,0,0.59,0.15}
\definecolor{verdes}{cmyk}{0.92,0,0.59,0.4}

\AtBeginDocument{
\heavyrulewidth=.08em
\lightrulewidth=.05em
\cmidrulewidth=.03em
\belowrulesep=.65ex
\belowbottomsep=0pt
\aboverulesep=.4ex
\abovetopsep=0pt
\cmidrulesep=\doublerulesep
\cmidrulekern=.5em
\defaultaddspace=.5em
}

\hypersetup{
 colorlinks=true,
 linkcolor=verdec,
 urlcolor=verdec,
 citecolor=verdec%gray
}

\newcommand{\lf}{\left(}
\newcommand{\ri}{\right)}

\newcommand{\nn}{\nonumber}

\newcommand{\hf}{h_{\rm f}}
\newcommand{\hfi}{h_{\rm f}^{(i)}}
\newcommand{\hfo}{h_{\rm f}^{(1)}}
\newcommand{\hft}{h_{\rm f}^{(2)}}
\newcommand{\rcol}{\gamma_{\rm col}}
\newcommand{\egmj}{J^\gamma_{\rm HCAL}}
\newcommand{\lecal}{L_{\rm ECAL}}
\newcommand{\lhcali}{L_{\rm HCAL}^i}
\newcommand{\lhcalf}{L_{\rm HCAL}^f}

\newcommand{\rr}{{\gamma\gamma}}

\newcommand{\rhad}{R_{\rm had}}

\newcommand{\hc}{{\rm H.c.}}

\newcommand{\fb}{{\;{\rm fb}}}

\newcommand{\iab}{{\;{\rm ab}^{-1}}}

\newcommand{\gev}{{\;{\rm GeV}}}
\newcommand{\tev}{{\;{\rm TeV}}}

\newcommand{\nsg}{N_{\rm S}}
\newcommand{\nbg}{N_{\rm B}}
\newcommand{\dbg}{\delta_{\rm B}}
\newcommand{\Dbg}{\Delta_{\rm B}}

\newcommand{\beq}{\begin{equation}}
\newcommand{\eeq}{\end{equation}}
\newcommand{\bea}{\begin{eqnarray}}
\newcommand{\eea}{\end{eqnarray}}
\newcommand{\barr}{\begin{array}}
\newcommand{\earr}{\end{array}}
\newcommand{\bc}{\begin{center}}
\newcommand{\ec}{\end{center}}
\newcommand{\bit}{\begin{itemize}}
\newcommand{\eit}{\end{itemize}}
\newcommand{\ben}{\begin{enumerate}}
\newcommand{\een}{\end{enumerate}}

\newcommand{\al}{\alpha}

\newcommand{\Dt}{\Delta}

\newcommand{\sg}{\sigma}

\newcommand{\gm}{\gamma}

\newcommand{\Gm}{\Gamma}
\newcommand{\lm}{\lambda}

\newcommand{\lmc}{\Lambda_{\rm cut}}

\newcommand{\drad}{d_{\rm rad}}
\newcommand{\hsm}{{h_{\rm SM}}}
\newcommand{\ch}{H^\pm}

\newcommand{\wpm}{W^\pm}

\newcommand{\mch}{M_{H^\pm}}
\newcommand{\mhh}{M_{H}}
\newcommand{\ma}{M_{A}}
\newcommand{\mach}{M_{A/H^\pm}}
\newcommand{\mhf}{m_{h_{\rm f}}}
\newcommand{\mbsq}{\overline{M}^2}
\newcommand{\mb}{\overline{M}}

\newcommand{\delphes}{{\small\sc Delphes}~}
\newcommand{\mdf}{{\small\sc MadGraph5\_aMC@NLO}}
\newcommand{\nnpdf}{{\small\sc NNPDF31\_nlo\_as\_0118}}

\newcommand{\tb}{t_\beta}

\newcommand{\cb}{c_\beta}
\renewcommand{\sb}{s_\beta}

\newcommand{\cba}{c_{\beta-\alpha}}
\newcommand{\sba}{s_{\beta-\alpha}}

\newcommand{\met}      {{E_T^{\rm miss}}}

\newcommand{\pt}      {p_T}

\newcommand{\ca}      {c_\alpha}
\newcommand{\sa}      {s_\alpha}

\definecolor{mint}{rgb}{0.24, 0.71, 0.54}

\begin{document}
\preprint{KIAS-P25016}

\title{\color{verdes} Emerging Photon Jets in the Hadronic Calorimeter: \\
A Novel Signature of Neutral Long-Lived Particles at the LHC
} 
\author{Jinheung Kim}
\email{jinheung.kim1216@gmail.com}
\address{School of Physics, Korea Institute for Advanced Study, 85 Hoegi-ro, Dongdaemun-gu, Seoul 02455, Republic of Korea} 
\author{Dongjoo Kim}
\email{dongzoo5260@gmail.com}
\address{Department of Physics, Konkuk University, 120 Neungdong-ro, Gwangjin-gu, Seoul 05029, Republic of Korea} 
\author{Soojin Lee}
\email{soojinlee957@gmail.com}
\address{Department of Physics, Konkuk University, 120 Neungdong-ro, Gwangjin-gu, Seoul 05029, Republic of Korea} 
\author{Jeonghyeon Song}
\email{jhsong@konkuk.ac.kr}
\address{Department of Physics, Konkuk University, 120 Neungdong-ro, Gwangjin-gu, Seoul 05029, Republic of Korea} 

\begin{abstract} 
We propose a novel collider signature for neutral long-lived particles (LLPs): the emerging photon jet in the hadronic calorimeter (HCAL). This signature arises when a neutral LLP decays into photons within the HCAL, producing an electromagnetic shower without associated charged tracks or energy deposits in the electromagnetic calorimeter (ECAL). To demonstrate the viability of this approach, we consider the fermiophobic Higgs boson $h_{\rm f}$ in the Type-I two-Higgs-doublet model as a representative scenario. In the ultralight regime ($m_{h_{\rm f}} < 1$~GeV), $h_{\rm f}$ decays exclusively into a photon pair via loop-induced processes, resulting in a suppressed width and consequently a long lifetime. Focusing on the golden channel $pp \to H^\pm h_{\rm f} \to W^\pm h_{\rm f} h_{\rm f}$, we analyze the exotic final state in which one $h_{\rm f}$ decays in the ECAL and appears as a highly collimated photon jet (reconstructed as a single photon), while the other decays within the HCAL, producing an emerging photon jet. Through a detailed signal-to-background analysis incorporating realistic detector effects via fast simulation, we demonstrate that this signature achieves discovery-level sensitivity at the HL-LHC across a broad region of parameter space consistent with theoretical and experimental constraints. While our study focuses on the fermiophobic Higgs, the emerging photon jet in the HCAL constitutes a broadly applicable and previously unexplored strategy for detecting neutral LLPs decaying into photons, opening a new avenue in LLP searches at colliders.
\end{abstract}
\vspace{1cm}

\maketitle
\tableofcontents
\flushbottom 

\section{Introduction}

The discovery of the Higgs boson at the LHC~\cite{ATLAS:2012yve,CMS:2012qbp} completed the Standard Model (SM), yet several fundamental questions remain unresolved—including the hierarchy problem, the origin of fermion mass hierarchies, the matter-antimatter asymmetry, the nature of neutrino masses, and the identity of dark matter. These open puzzles strongly motivate the search for new physics beyond the SM (BSM).
Among the compelling candidates predicted by BSM theories are long-lived particles (LLPs)~\cite{Curtin:2018mvb,Alimena:2019zri}, which appear naturally in a broad class of well-motivated scenarios, including supersymmetry~\cite{Giudice:1998bp,Evans:2013kxa,Ambrosanio:1996jn,Chemtob:2004xr,Barbier:2004ez,Meade:2008wd,Meade:2010ji,Knapen:2016exe}, axion-like particles~\cite{Goto:1991gq,Chun:1992zk}, Neutral Naturalness~\cite{Chacko:2005pe,Burdman:2006tz,Cai:2008au,Craig:2014aea}, composite Higgs models~\cite{Panico:2015jxa}, and dark sector frameworks~\cite{Hall:2009bx}.

From a phenomenological perspective, LLPs have acquired increasing attention as promising discovery channels---particularly in light of null results from conventional BSM searches at high-energy colliders. Because standard search strategies typically assume prompt decays near the interaction point, LLP signatures can escape detection by traditional triggers and reconstruction algorithms. Charged LLPs are generally easier to identify, as they often leave anomalously large ionization energy loss ($dE/dx$) signals while traversing the detector~\cite{ATLAS:2015wsk,ATLAS:2016onr,CMS:2016kce,ATLAS:2018lob,ATLAS:2018imb,ATLAS:2019gqq,ATLAS:2022pib,ATLAS:2023zxo,Vami:2024ocv,CMS:2024nhn}.

In contrast, neutral LLPs---lacking direct interactions with the tracking material---pose a greater experimental challenge and require more specialized detection strategies. A common approach involves searching for displaced vertices with large impact parameters~\cite{Guo:2021vpb,Calibbi:2021fld}. Numerous experimental signatures have been explored at the LHC, including displaced vertices in the inner tracking detector~\cite{ATLAS:2019jcm,ATLAS:2020xyo,ATLAS:2021jig,ATLAS:2024qoo}, 
displaced muons~\cite{ATLAS:2018rjc,Yuan:2020eeu,CMS:2021juv,CMS:2021sch,ElFaham:2022vot,CMS:2022qej,Santocchia:2024ayt,CMS:2024bvl,CMS:2024qxz,Ruiz:2025akn}, 
displaced jets~\cite{ATLAS:2018niw,CMS:2018tuo,Bhattacherjee:2019fpt,Chiang:2019ajm,CMS:2018qxv,CidVidal:2019urm,ATLAS:2019qrr,Liu:2020vur,CMS:2020iwv,CMS:2021yhb,CMS:2021tkn,ATLAS:2022gbw,ATLAS:2022zhj,Carmona:2022jid,Kucharczyk:2022pie,CMS:2022wjc,Bhattacherjee:2023kxw,ATLAS:2023oti,ATLAS:2024ocv,CMS:2024xzb}, displaced jets with missing transverse energy~\cite{ATLAS:2017tny,CMS:2019qjk,Bernreuther:2020xus,ATLAS:2023cjw,CMS:2024trg}, displaced leptons~\cite{ATLAS:2015oan,Evans:2016zau,Jones-Perez:2019plk,Liu:2019ayx,ATLAS:2019fwx,ATLAS:2019tkk,ATLAS:2020wjh,CMS:2022fut}, displaced $\ell^\pm j$~\cite{CMS:2023jqi,CMS:2024hik}, displaced photons in the electromagnetic calorimeter (ECAL)~\cite{Kawagoe:2003jv,CMS:2012bbi,ATLAS:2023ian}, out-of-time calorimeter deposits~\cite{ATLAS:2021mdj}, and $\mu W$ signals in the muon system~\cite{CMS:2024ake}.\footnote{Complementary studies have been conducted at LHCb~\cite{LHCb:2016inz,LHCb:2017xxn,Gligorov:2017nwh,LHCb:2020akw,LHCb:2021dyu}, Belle II~\cite{Filimonova:2019tuy,Duerr:2020muu,Acevedo:2021wiq,Zhou:2021ylt,Ferber:2022ewf,Ferber:2022rsf}, future $e^+e^-$ colliders~\cite{Schafer:2022shi,Ripellino:2024tqm}, and Lifetime Frontier experiments~\cite{Berlin:2018jbm,Alimena:2019zri,Acharya:2020uwc,MATHUSLA:2020uve,Araki:2020wkq,NA62:2020xlg,DeVries:2020jbs,Du:2021cmt,Anchordoqui:2021ghd,Cottin:2021lzz,MicroBooNE:2022ctm,Kyselov:2024dmi,Lezki:2024ueg}, as well as at DUNE~\cite{CMS:2021tkn}, Super-Kamiokande~\cite{Arguelles:2019ziu,Candia:2021bsl}, and future $ep$ colliders~\cite{Cheung:2020ndx,Gu:2022muc}.}

Despite the extensive literature on LLP signatures, one important possibility has remained unexplored: the decay of a neutral LLP into photons inside the hadronic calorimeter (HCAL). The resulting electromagnetic shower initiates deep within the HCAL, producing a distinctive signal without associated charged tracks or energy deposits in the ECAL. Unlike typical QCD jets---which deposit a substantial fraction of their energy in the ECAL due to prompt neutral pion decays---this emerging jet is both neutral and spatially displaced. We refer to this novel signature as the emerging photon jet in the HCAL, denoted by $\egmj$.

To demonstrate the potential of this new strategy, we consider a concrete BSM framework: the Type-I two-Higgs-doublet model (2HDM) featuring an ultralight fermiophobic Higgs boson. In the inverted Higgs scenario, where the heavier  \textit{CP}-even scalar $H$ is identified as the 125 GeV Higgs boson observed at the LHC, the condition $\alpha = \pi/2$ with $\alpha$ being the mixing angle between the  \textit{CP}-even states $h$ and $H$ renders the lighter scalar $h$ fermiophobic, denoted by $\hf$. When $\hf$ is ultralight, with $m_{\hf} < 1$ GeV, it decays exclusively to a photon pair via loop-induced processes, resulting in a suppressed decay width and hence a long lifetime. We perform a detailed scan of the model's parameter space, demonstrating that theoretical and experimental constraints impose an upper bound on the charged Higgs mass, $\mch \lesssim 335$ GeV. Since the charged Higgs decays almost exclusively to $W^\pm \hf$, this relatively light $\mch$ makes the associated production process $pp \to \ch \hf \to W^\pm \hf \hf$ accessible at the LHC across the full allowed parameter space. We identify this as the golden channel for probing $\hf$. Focusing on the exotic final state in which one $\hf$ decays in the ECAL, yielding a highly collimated photon pair reconstructed as a single photon, and the other decays within the HCAL, producing an $\egmj$, we perform a full signal-to-background analysis incorporating detector-level effects.

Our results showcase that this novel signature enables discovery-level sensitivity at the HL-LHC across a broad region of the allowed parameter space. While our analysis focuses on the fermiophobic Higgs scenario, the proposed signature is broadly applicable to any neutral LLP that dominantly decays into photons. To the best of our knowledge, this is the first dedicated study of emerging photon jets in the HCAL, opening a promising new avenue in the search for neutral LLPs at collider experiments.

\section{Review of the fermiophobic Higgs boson in Type-I 2HDM}
\label{sec-review}

The Two-Higgs-Doublet Model (2HDM) contains two $SU(2)_L$ complex scalar doublet fields with hypercharge $Y=1$ in the $Q = T_3 + Y$ convention~\cite{Branco:2011iw}:
\bea
\label{eq:phi:fields}
\Phi_i = \left( \begin{array}{c} w_i^+ \\[3pt]
\dfrac{v_i + \rho_i + i \eta_i}{\sqrt{2}}
\end{array} \right), \quad (i=1,2),
\eea
where $v_{1}$ and $v_2$ are the vacuum expectation values of $\Phi_{1}$ and $\Phi_2$, respectively. Their combination, $v =\sqrt{v_1^2+v_2^2}\approx 246\gev$, induces spontaneous electroweak symmetry breaking. A key parameter of the model is $\tb = v_2/v_1$,  where we define $s_x =\sin x$, $c_x = \cos x$, and $t_x = \tan x$, for notational simplicity.

To prevent flavor-changing neutral currents (FCNC) at tree level, we impose a discrete $Z_2$ symmetry under which $\Phi_1 \to \Phi_1$ and $\Phi_2 \to -\Phi_2$~\cite{Glashow:1976nt,Paschos:1976ay}. The scalar potential, respecting a softly broken $Z_2$ symmetry and \textit{CP} invariance, is given by
\begin{align}
\label{eq:VH}
V_\Phi &=  m^2 _{11} \Phi^\dagger_1 \Phi_1 + m^2 _{22} \Phi^\dagger_2 \Phi_2 - m^2 _{12} ( \Phi^\dagger_1 \Phi_2 + \hc) \nonumber\\[2mm]
&\quad + \frac{1}{2}\lambda_1 (\Phi^\dagger_1 \Phi_1)^2 + \frac{1}{2}\lambda_2 (\Phi^\dagger_2 \Phi_2)^2 + \lambda_3 (\Phi^\dagger_1 \Phi_1) (\Phi^\dagger_2 \Phi_2) \nonumber\\[2mm]
&\quad + \lambda_4 (\Phi^\dagger_1 \Phi_2)(\Phi^\dagger_2 \Phi_1) + \frac{1}{2} \lambda_5 \left[ (\Phi^\dagger_1 \Phi_2)^2 + \hc \right],
\end{align}
where the $m_{12}^2$ term softly breaks the $Z_2$ symmetry. This potential yields five physical Higgs bosons: the lighter \textit{CP}-even scalar $h$, the heavier \textit{CP}-even scalar $H$, the \textit{CP}-odd pseudoscalar $A$, and charged Higgs bosons $H^\pm$. For detailed relationships between mass eigenstates and weak eigenstates, see Ref.~\cite{Song:2019aav}, which introduces mixing angles $\alpha$ and $\beta$.

The SM Higgs boson $\hsm$ is a linear combination of $h$ and $H$, via $\hsm=\sba h+\cba H$. To ensure a long lifetime for the fermiophobic Higgs boson, we consider the ultralight mass regime below 1 GeV, which suppresses the decay width. This motivates our focus on the inverted scenario where the heavier \textit{CP}-even scalar $H$ corresponds to the observed Higgs boson with $\mhh = 125\gev$. In this setup, the Higgs coupling modifiers to the electroweak gauge bosons ($V = W^\pm, Z$) are given by $\kappa_V^H = \cba$ and $\xi_V^h = \sba$. Thus the SM-like behavior of the observed Higgs boson at the LHC requires $|\sba| \ll 1$.

In the Type-I 2HDM, the Yukawa coupling modifiers are $\kappa^H_f=\sa/\sb$, $\xi^h_f=\ca/\sb$, and $\xi^A_t=-\xi^A_{b,\tau}={1}/\tb$. Consequently, setting $\alpha=\pi/2$ renders $h$ fermiophobic,\footnote{This fermiophobic condition is preserved at loop level by an appropriate renormalization condition~\cite{Barroso:1999bf,Brucher:1999tx}.} denoted by $\hf$. To ensure compatibility with the Peskin–Takeuchi oblique parameters~\cite{Peskin:1991sw}, we further assume degenerate masses for $\ch$ and $A$. Our fermiophobic Type-I model is thus defined by
\bea\label{eq:setup}
\hbox{Type-I: }\mhh=125\gev,\quad\alpha=\frac{\pi}{2},\quad\mch=\ma\equiv\mach,
\eea
resulting in four model parameters: $\{\tb, m_{12}^2, \mhf, \mach \}$.

In this setup, $\tb$ is determined by $\sba$ through
\beq
\label{eq-tb}
\tb=-\frac{\cba}{\sba},
\eeq
which yields $|\sba|\simeq1/\tb$ for $\tb\gg1$. Exact Higgs alignment is thus excluded since it would cause divergent $\tb$. However, we consider approximate alignment $|\sba|\ll1$, requiring $\tb\gg1$.

With the condition $\al=\pi/2$, the quartic couplings in \autoref{eq:VH} are~\cite{Das:2015mwa}
\bea
\label{eq:quartic:lm1}
\lm_1 &\simeq& \frac{1}{v^2}
\left[
\mhf^2 + \tb^2 \lf \mhf^2 - \mbsq \ri
\right],
\\[3pt] \label{eq:quartic:lm2}
\lm_2 &\simeq& \frac{1}{v^2} \lf \frac{1+\tb^2}{\tb^2} \ri
\left[
\mhh^2  - \frac{\mbsq}{1+\tb^2}
\right],
\\[3pt] \label{eq:quartic:lm3}
\lm_3 &\simeq& \frac{1}{v^2}
\left[
2\mach^2 -\mbsq
\right],
\\[3pt] \label{eq:quartic:lm45}
\lm_4 &\simeq& \lm_5 \simeq \frac{1}{v^2}
\left[
\mbsq-\mach^2
\right],
\eea
where $\mbsq = m_{12}^2/(\sb\cb)$.

We now discuss the parameter space allowed by theoretical requirements and experimental constraints. 
For a given $\mhf = 0.5\gev$, we perform a random scan over the following ranges: 
\bea \label{eq:scan:range} 
\mach &\in& \left[ 80, 600 \right] \gev, \quad \tb \in \left[3,50 \right], \quad m_{12}^2 \in \left[ 0, 1 \right] \gev^2. 
\eea 
The scanning range for $m_{12}^2$ is deliberately restricted, as preliminary analyses show that only positive values below $1\gev^2$ remain viable.

To determine the allowed parameter points, we first impose several theoretical requirements, including the bounded-from-below criterion for the Higgs potential~\cite{Ivanov:2006yq}, tree-level unitarity of scalar-scalar scattering processes~\cite{Branco:2011iw,Arhrib:2000is}, and perturbativity of Higgs quartic couplings~\cite{Chang:2015goa}. These constraints are evaluated using the public code \textsc{2HDMC}~\cite{Eriksson:2009ws}. Additionally, we enforce vacuum stability condition by requiring $D=\left[ m_{11}^2 - (\lm_1/\lm_2)^{1/2} m_{22}^2 \right]\left[\tb- (\lm_1/\lm_2)^{1/4} \right]>0$~\cite{Ivanov:2008cxa,Barroso:2012mj,Barroso:2013awa}.

We further ensure consistency with the latest best-fit values of the Peskin-Takeuchi electroweak oblique parameters~\cite{Peskin:1991sw} in the 2HDM framework~\cite{He:2001tp,Grimus:2008nb}: $S=-0.04 \pm 0.10$, $T= 0.01 \pm 0.12$, and $U=-0.01\pm 0.09$,
properly accounting for their correlations~\cite{ParticleDataGroup:2024cfk}. Finally, we require that the cutoff scale $\lmc$---the energy at which unitarity, perturbativity, or vacuum stability breaks down---exceeds $10\tev$~\cite{Kim:2023lxc}. This scale is determined using the public code \textsc{2HDME}~\cite{Oredsson:2018yho,Oredsson:2018vio}.
For low-energy experimental constraints, we incorporate FCNC data at the 95\% confidence level, 
including $B \to X_s \gamma$~\cite{Arbey:2017gmh,Sanyal:2019xcp,Misiak:2017bgg}, $B \to K^* \gamma$~\cite{Belle:2017hum}, and $B_s \to \phi \gamma$~\cite{Belle:2014sac}. The constraints from Higgs precision measurements and direct searches at the LEP, Tevatron, and LHC are also checked using the public code \textsc{HiggsTools}-v1.2~\cite{Bahl:2022igd}. 

Our parameter scan initially produced $10^6$ points that satisfy theoretical requirements and electroweak precision constraints. The application of FCNC constraints removed approximately 7.8\% of these. Additionally, enforcing the cutoff scale requirement $\lmc > 10\tev$ led to the exclusion of a further 53.3\%. After applying the constraints from Higgs precision measurements and direct search bounds, we found that about 28.5\% of the original parameter space remained viable.

\begin{figure}[!t]
\centering
\includegraphics[width=0.6\textwidth]{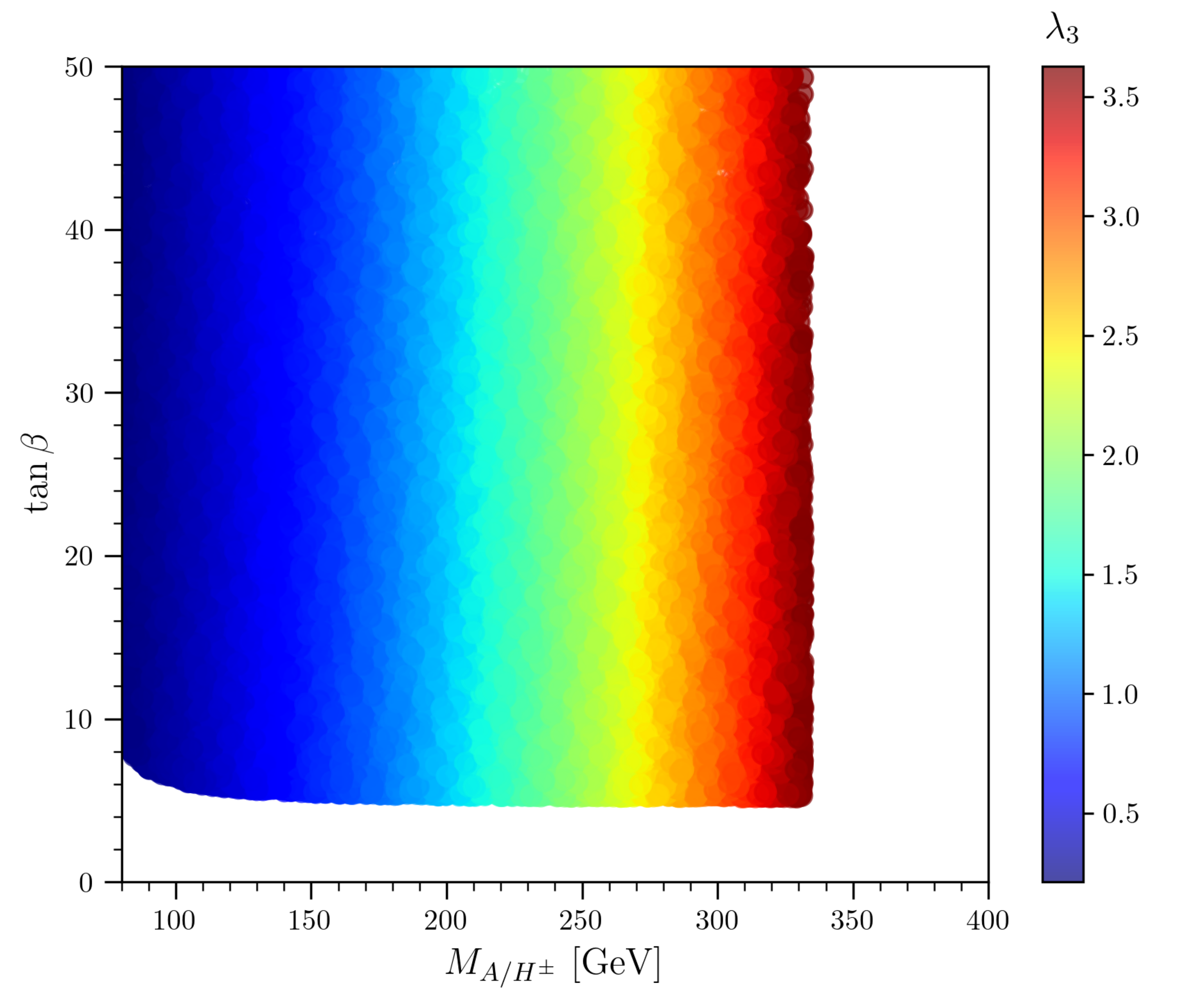} 
\vspace{-0.7cm}
\caption{Allowed parameter points in $(\mach,\tb)$ space with the color code for the quartic coupling $\lm_3$.
We set $\mhf=0.5\gev$ and $\al=\pi/2$. }
\label{fig-allowed}
\end{figure}

We now examine the key features of the allowed parameter space. One of the most striking aspects is the presence of a strict upper bound on $m_{12}^2$, which must be positive and lie below approximately $(0.215\gev)^2$. This strong limitation is primarily driven by the global vacuum stability condition. 

\autoref{fig-allowed} illustrates the distribution of the remaining model parameters, $\mach$ and $\tb$, with the color coding indicating the value of $\lambda_3$. Two notable features exist in the fermiophobic Type-I scenario. First, a lower bound appears on $\tb$ around 5.2, which becomes more stringent for lighter charged Higgs masses. For example, when $\mch = 80\gev$, the allowed region requires $\tb \gtrsim 8$. This behavior primarily stems from Higgs precision measurements, particularly in the process $gg \to H \to \gamma\gamma$. Since the charged Higgs boson contributes to this loop-induced process~\cite{CMS:2021kom}, a lighter $\mch$ necessitates a suppression of the top quark Yukawa coupling, which is achieved by increasing $\tb$.

Second, an upper bound is observed on the allowed $\mach$, restricting it to values below approximately $335\gev$. This suggests that the fermiophobic Type-I scenario disfavors very heavy BSM Higgs bosons, placing $A$ and $\ch$ within the LHC's kinematic reach. The primary driver of this upper bound on $\mach$ is the behavior of $\lm_3$, as indicated by the color code. A clear correlation exists between $\mach$ and $\lm_3$, where larger $\lm_3$ corresponds to heavier $\mach$, a relationship explicitly seen in \autoref{eq:quartic:lm3}. Within the allowed parameter space, the requirement $\mb \lesssim 0.52\gev$ effectively enforces $\lambda_3$ to scale proportionally with $\mach^2$. As excessively large values of $\lambda_3$ violate unitarity or perturbativity, this correlation places a strong upper limit on $\mach$.

Now let us examine the decay patterns of the BSM Higgs bosons. Across the entire allowed parameter space, both the charged Higgs boson and the pseudoscalar exhibit nearly exclusive decay modes: $\ch \to W^\pm \hf$ and $A \to Z\hf$. As a result, the collider signatures of the fermiophobic Type-I scenario  are largely dictated by the behavior of $\hf$. Since $\hf$ is fermiophobic and the lightest scalar in the model, it has only one viable decay mode, $\hf \to \gamma\gamma$, which proceeds via loop contributions from $W^\pm$ and $H^\pm$.

The partial decay width is given by~\cite{Djouadi:2005gj}
\bea
\label{eq-decay-width}
\Gm(\hf\to\rr) = \frac{\al^2 \mhf^3}{256 \pi^3 v^2}
\left|
\sba A^{\mathcal{H}}_1(\tau_W) + \frac{v^2}{2 \mch^2} \, \hat{g}_{\hf H^+ H^-}  \,  A^{\mathcal{H}}_0(\tau_{\ch})
\right|^2,
\eea
where $\tau_i = \mhf^2/(4 m_i^2)$
and the loop functions $ A^{\mathcal{H}}_{1,0}(x)$ are referenced in Ref.~\cite{Djouadi:2005gj}.
The normalized coupling $\hat{g}_{\hf H^+ H^-} $ is given by~\cite{Bernon:2015qea}
\beq
\label{eq-g-hH+H-}
\hat{g}_{\hf H^+ H^-} = -\frac{1}{v^2}
\left[
\sba \left\{ \mhf^2 + 2 \lf \mch^2-\mbsq\ri \right\} + \cba\, \frac{1-\tb^2}{\tb}
\lf \mhf^2-\mbsq\ri
\right].
\eeq
Since the partial width $\Gamma(\hf \to \gamma\gamma)$ scales as $\mhf^3$, an ultralight $\hf$ leads to a highly suppressed decay rate. As a result, the decay length becomes sizable, rendering $\hf$ a potential LLP.

\begin{figure}[!t]
\centering
\includegraphics[width=0.55\textwidth]{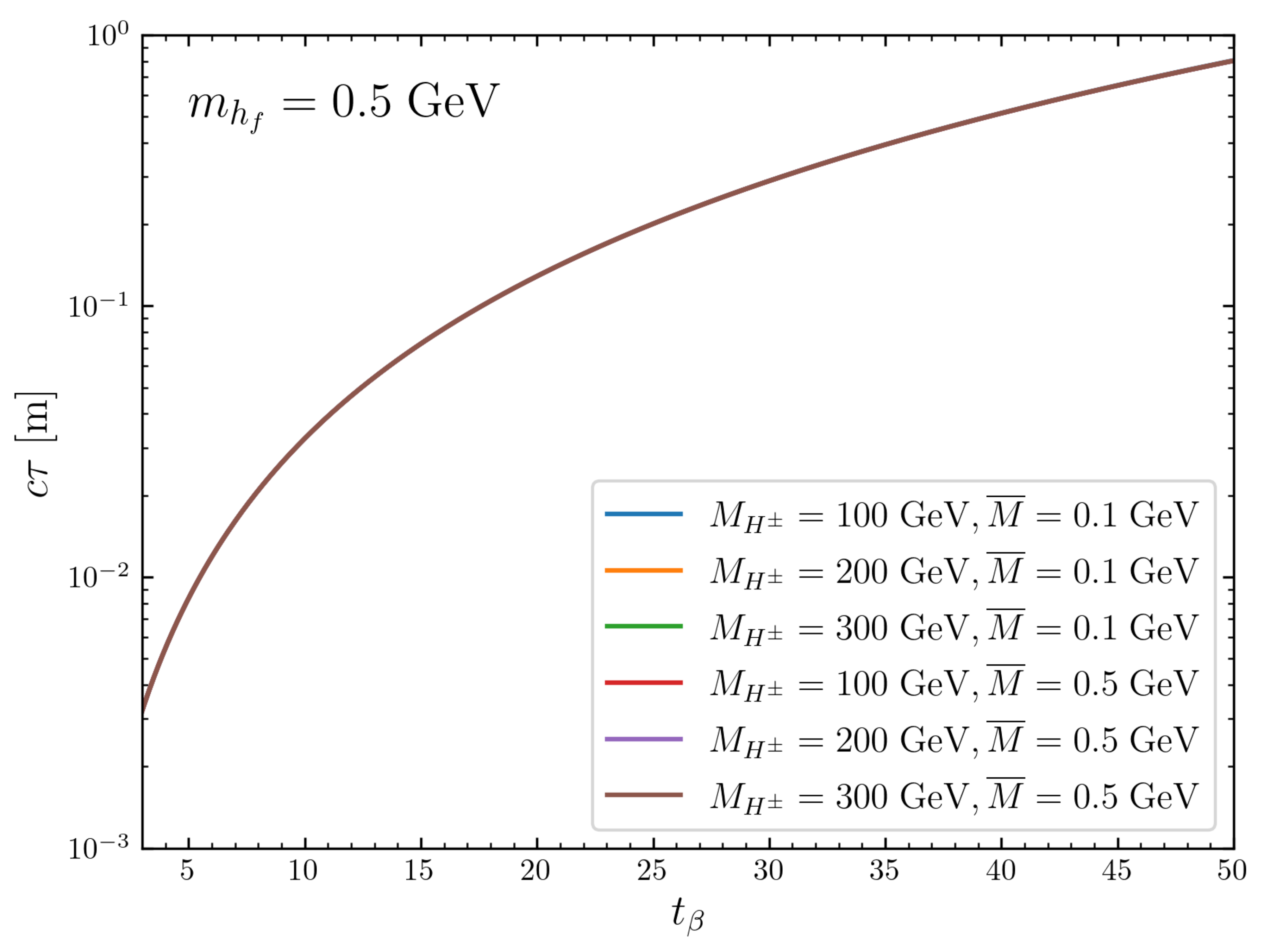} 
%\vspace{-0.7cm}
\caption{Decay length $c\tau$ in the rest frame of $\hf$, as a function of $\tb$. For a given $\mhf=0.5\gev$, we consider six cases of $(\mch,\mb)$, where $\mch=100,200,300\gev$ and $\mb=0.1,0.5\gev$.
 }
\label{fig-ctau}
\end{figure}

In \autoref{fig-ctau}, we present the Lorentz-invariant decay length $c\tau$ of $\hf$ as a function of $\tb$, for benchmark scenarios with $\mach = 100, 200, 300\gev$ and $\mb = 0.1, 0.5\gev$. Remarkably, all cases follow a single universal curve, indicating that $c\tau$ is practically independent of these mass parameters. This behavior originates from the structure of $\hat{g}_{\hf H^+ H^-}$ in \autoref{eq-g-hH+H-}. For an ultralight $\hf$ with mass below 1 GeV, theoretical and experimental constraints require $\mb \lesssim 0.52\gev$, under which the term proportional to $\sba \mch^2$ dominates.\footnote{The term proportional to $\cba$ in $\hat{g}_{\hf H^+ H^-}$ can become dominant if $|\sba| \lesssim 10^{-3}$. However, this corresponds to an extremely large $\tb \gtrsim 10^3$, a highly fine-tuned regime that we do not consider.} As a result, the $\mch$ dependence in the second term of \autoref{eq-decay-width} effectively cancels. The partial width $\Gamma(\hf \to \gamma\gamma)$ then becomes proportional to $\sba^2$, which in turn scales as $1/\tb^2$ via the relation in \autoref{eq-tb}. This explains the observed increase in $c\tau$ with growing $\tb$.

The most notable feature of the results in \autoref{fig-ctau} is the sizable proper decay length $c\tau$ of the ultralight $\hf$, ranging from $\mathcal{O}(1)$ mm to $\mathcal{O}(1)$ m. This highlights the potential for $\hf$ to manifest as a long-lived particle (LLP). Experimentally, the observable decay length in the laboratory frame is given by $c\tau\beta\gamma$, where $\beta\gamma = |\vec{p}|/\mhf$. However, at the LHC, the longitudinal momentum of the partonic center-of-mass frame is unknown on an event-by-event basis, making it impossible to precisely reconstruct the total momentum $|\vec{p}|$ of the LLP. Combined with the cylindrical geometry of the ATLAS and CMS detectors, this limitation motivates LLP analyses to focus on the transverse displacement from the beamline. As a result, the transverse decay radius $\drad = c\tau \beta_T \gamma$ serves as the most practical and experimentally relevant observable. Given that $\beta_T \gamma = p_T/\mhf$, the ultralight nature of $\hf$ implies that $\drad$ can greatly exceed the proper decay length $c\tau$. 
 
Depending on where $\hf$ decays, different detector signatures arise. If  $\hf$ decays into $\gamma\gamma$ before reaching the outer boundary of the ECAL, the resulting highly collimated photon pair is most often reconstructed as a single photon.\footnote{If $\mhf$ lies between 1 GeV and 10 GeV, the photons are close enough to form a jet but sufficiently separated to exhibit diphoton-like substructure, forming a diphoton jet. Although these photons produce a single localized energy deposit in the ECAL, mimicking a single-photon jet, a Convolutional Neural Network (CNN)-based analysis demonstrated in Refs.~\cite{Ren:2021prq,Wang:2023pqx} can distinguish this diphoton-jet signature from QCD jets and isolated photons.} due to their small angular separation, approximately given by $R_{\gamma\gamma} \sim 2 \mhf / \pt$~\cite{Ren:2021prq,Wang:2023pqx}. For instance, with $\mhf = 0.5\gev$ and $p_T = 50\gev$, the separation is $\Delta R \sim 0.02$, which is smaller than the minimum granularity of the ECAL, $R_{\rr}^{\rm min} = 0.035$~\cite{delPeso:2011ica}. We refer to this highly collimated photon jet, reconstructed as a single photon in the ECAL, as $\rcol$.

A more exotic signature arises when $\hf$ decays within the HCAL volume. In this case, the resulting photon pair interacts directly with the calorimeter material, producing electromagnetic showers inside the HCAL. These showers are reconstructed as a jet. Since there are no associated charged tracks in the inner detector, the jet is neutral. As this photon-induced jet appears to emerge from within the HCAL, we refer to this distinctive signature as an emerging photon jet in the HCAL, denoted by $\egmj$.

If $\hf$ decays within the muon spectrometer (MS), its experimental signature becomes more difficult to analyze. Since both the ATLAS and CMS MS systems rely on charged-particle ionization and magnetic deflection, neutral photons typically traverse the spectrometer without detection, contributing to missing transverse energy.\footnote{When a LLP decays into hadronic jets or a charged lepton pait, their displaced vertices can be detecter~\cite{Bondarenko:2019tss,CMS:2021juv,ATLAS:2022gbw}.} However, photons can occasionally convert into electron-positron pairs through interactions with detector material. These converted $e^+e^-$ pairs would produce signatures distinct from typical muons, most notably lacking associated tracks in the inner detector. Given that the CMS MS includes an iron return yoke, the photon conversion rate is expected to be significantly higher in CMS than in ATLAS. A dedicated study of this photon-conversion-induced signature lies beyond the scope of this letter. Finally, if $\hf$ decays beyond the MS, its signal consists solely of missing transverse energy.

Among these  possible signatures of $\hf$ at the LHC, we focus on $\rcol$ and $\egmj$. These complementary signatures capture the characteristic phenomenology of long-lived $\hf$ decays and form the foundation of our proposed search strategy.

 \section{Signal-to-Background analysis for $pp \to \ch \hf \to \wpm \hf\hf$}
 \label{sec-analysis}

In this section, we quantitatively evaluate the LHC discovery potential for $\hf$ by performing a full signal-to-background analysis at the detector level.
A crucial step  is identifying a promising production channel for $\hf$. Since $\hf$ is fermiophobic, it cannot be produced via gluon fusion through quark loops. Instead, the dominant production modes involve associated production with other BSM Higgs bosons,  $pp \to W^* \to \hf\ch$ and $pp \to Z^* \to \hf A$~\cite{Kim:2022nmm,Kim:2023lxc}. Among these, the $W$-mediated process is  favorable, as it includes both charge-conjugated subprocesses.

The production cross sections for the signal process $pp \to \hf \ch$ are sizable at the 14 TeV LHC. For benchmark cases with $\mch = 100, 200, 300\gev$ and fixed $\tb = 10$ (corresponding to $\cba = 0.995$ in the fermiophobic Type-I scenario), we compute the parton-level cross sections using \mdf~\cite{Alwall:2011uj} version 3.5.6 with the \nnpdf PDF set: 
\bea \label{eq-tot-xs-signal} 
\sigma(pp \to \hf\ch )\big|_{14\tev}
=
\left\{ 
\begin{array}{ll} 
2.79\times 10^3\fb & \hbox{if } \mch = 100\gev, \\
 2.18\times 10^2\fb & \hbox{if } \mch = 200\gev, \\
  5.20\times 10^1\fb & \hbox{if } \mch = 300\gev. 
\end{array} \right. 
\eea 
We observe that the signal cross section decreases rapidly with increasing charged Higgs boson mass, illustrating the strong sensitivity of the production rate to $\mch$. 
Although the above calculation assumes a fixed $\tb$, we find that the dependence on $\tb$ is relatively mild. Since the cross section scales with $\cba^2$, its $\tb$ dependence enters only indirectly through the relation in \autoref{eq-tb}. Given that $\cba$ is tightly constrained to be above 0.95 by Higgs precision data, variations in $\tb$ within the allowed range induce only minor changes in the production rate.

We now turn to the final-state topology of the signal process $pp \to \hf \ch$. Given that $\ch \to \hf W^\pm$ proceeds with nearly 100\% branching ratio, the process yields two $\hf$ bosons in the final state. We focus on the scenario in which one $\hf$ decays within the ECAL and is reconstructed as $\rcol$, while the other decays inside the HCAL and gives rise to an $\egmj$. This leads to our golden channel: 
\bea \label{eq-signal-process} 
pp \to \ch \hf \to \wpm \hf\hf \to \ell^\pm \nu \, \rcol \, \egmj,
\eea
 where $\ell^\pm = e^\pm, \mu^\pm$. The resulting final state consists of a charged lepton, missing transverse energy ($\met$), a single photon, and an emerging photon jet in the HCAL.

 \begin{figure}[!t]
\centering
\includegraphics[width=0.65\textwidth]{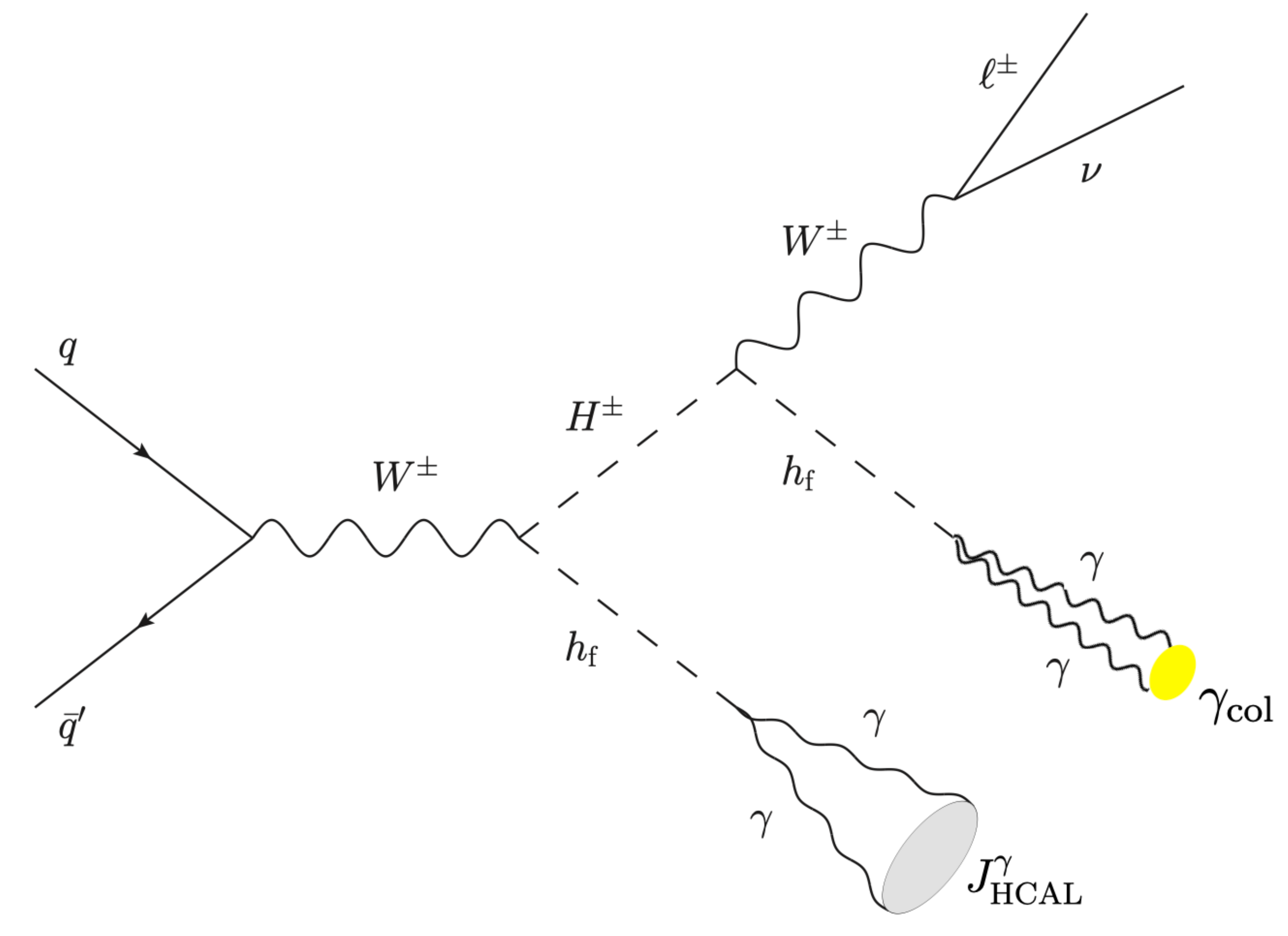} 
\caption{Representative Feynman diagram for the signal process $pp \to \ch \hf \to W^\pm \hf\hf \to \ell^\pm \nu \, \rcol \, \egmj$. In this illustration, the $\hf$ produced directly in the scattering process is detected as an emerging photon jet in the HCAL ($\egmj$), while the $\hf$ from the decay of the charged Higgs boson $\ch$ is reconstructed as a single photon jet in the ECAL ($\rcol$). The roles of the two $\hf$ bosons are interchangeable; the prompt $\hf$ can also be detected as $\rcol$, with the decay product of $\ch$ appearing as $\egmj$.
}
\label{fig-Feynman}
\end{figure}

In \autoref{fig-Feynman}, we present a representative Feynman diagram for this signal process, where the $\hf$ produced directly in the scattering process is reconstructed as $\egmj$, and the $\hf$ from the decay of the charged Higgs boson is reconstructed as $\rcol$. We note, however, that the roles of the two $\hf$ bosons are interchangeable: the prompt $\hf$ can also be detected as $\rcol$, while the $\hf$ from the $\ch$ decay appears as $\egmj$.

For the background, we consider the $2\to4$ process that directly matches the signal final state: 
\beq
\label{eq-BG}
pp \to \ell^\pm \nu j \gm.
\eeq
This background includes both the hard scattering process $pp \to W^\pm (\to \ell^\pm \nu) j \gamma$ and the process $pp \to \ell^\pm \nu j$ with an additional photon emitted via final-state bremsstrahlung.\footnote{Among bremsstrahlung sources, radiation from electrons is particularly significant due to their small mass, which enhances collinear photon emission~\cite{Karliner:2015tga}. Photon bremsstrahlung from muons is generally suppressed due to their larger mass, and photons radiated from quarks are also less relevant, as they tend to be clustered with the jet.
} Although the photon origins differ, the resulting final states are experimentally indistinguishable and must be treated as a single irreducible background in the analysis.

A comprehensive workflow was employed to simulate both signal and background events. The signal samples correspond to three benchmark scenarios with $\mch = 100,\ 200,\ 300\gev$, assuming fixed values of $\tb = 10$ and $\mhf = 0.5\gev$.
Parton-level cross sections at the 14 TeV LHC were calculated using \mdf~version 3.5.6, with the \nnpdf PDF set. The renormalization and factorization scales were set to $\mu_R = \mu_F = \sum_i \sqrt{p_{T,i}^2 + m_i^2}$. We generated $2 \times 10^5$ signal events for each benchmark case and $2.5 \times 10^6$ background events for the background process. To improve generation efficiency and suppress soft bremsstrahlung contributions, we imposed a generator-level cut of $\pt^{j,\gamma} > 40\gev$ for the background events, which is justified since we later apply tighter $\pt$ cuts in the event selection.

Initial-state radiation, parton showering, and hadronization were performed using {\small\sc Pythia} version 8.309~\cite{Bierlich:2022pfr}. Detector response was modeled using \delphes  version 3.5.0~\cite{deFavereau:2013fsa} with the high-luminosity LHC configuration card \texttt{delphes\_card\_HLLHC.tcl}. Jet reconstruction was carried out using {\small\sc FastJet} version 3.3.4~\cite{Cacciari:2011ma} with the anti-$k_T$ algorithm and a jet radius parameter of $R = 0.4$. For photon isolation, we adopted the default definition in the \delphes card: a photon is considered isolated if the scalar sum of transverse momenta of all particles within a cone of $\Delta R < 0.3$ around the photon, excluding the photon itself, is less than 20\% of the photon's $\pt$~\cite{deFavereau:2013fsa,Martins:2022dfg}.

For the signal process with final-state particles as in \autoref{eq-signal-process}, we compute the probability that one $\hf$ is detected as $\rcol$ and the other as $ \egmj$. To be detected as $\rcol$, $\hf$ must decay before reaching the outer boundary of the ECAL. Conversely, for $\hf$ to be reconstructed as $\egmj$, it must decay within the HCAL volume. These probabilities follow exponential decay statistics, governed by the decay radius, which in turn depends on the transverse momentum of each $\hf$.

Let us consider a signal event in which the two $\hf$ bosons have transverse momenta $p_{T}^{(1)}$ and $p_{T}^{(2)}$, denoted by $\hfo$ and $\hft$, respectively. The decay radius for each $\hf^{(i)}$ is given by $\drad^{(i)} = c\tau p_T^{(i)} / \mhf$. The probabilities that $\hf^{(i)}$ is reconstructed as $\rcol$ or $\egmj$ are then expressed as
\begin{align} \label{eq-prob} 
P(\hfi\to \rcol) &= 1 - \exp \left( - \frac{\lecal}{\drad^{(i)}} \right), \\ \nn 
P(\hfi\to \egmj) &= \exp \left( - \frac{\lhcali}{\drad^{(i)}} \right) - \exp \left( - \frac{\lhcalf}{\drad^{(i)}} \right), \end{align} 
where $\lecal$ is the radial distance to the outer boundary of the ECAL, $\lhcali$ marks the inner boundary of the HCAL, and $\lhcalf$ its outer boundary.

The ATLAS and CMS detectors have different geometrical dimensions for their ECAL and HCAL. Since ATLAS has larger ECAL and HCAL volumes, it provides a greater geometric acceptance for decays occurring within the HCAL. Thus, we adopt the ATLAS detector configuration in our analysis, taking $\lecal = 2.0 \;{\rm m}$, $\lhcali = 2.25 \;{\rm m}$, and $\lhcalf = 3.9 \;{\rm m}$~\cite{Henriques:2015fso,ATLAS:2018edp}.

For both signal and background events at the detector level, we apply the following basic selection criteria: \begin{itemize}
\item At least one photon, one jet, and one lepton are required: 
$N_\gamma \geq 1$, $N_j \geq 1$, and $N_\ell \geq 1$, 
with all objects satisfying the pseudorapidity condition $|\eta| < 2.5$. 
For the leading jet and leading photon (ordered by descending $p_T$), 
we impose a transverse momentum requirement of $\pt^{j_1,\gamma_1} > 70\gev$. 
\item Missing transverse energy is required to satisfy $E_T^{\rm miss} > 50\gev$. 
\end{itemize}

To design an effective cut-flow strategy for enhancing the signal significance, we examined various kinematic variables that can discriminate the signal from the background. Among them, we identified four powerful variables:
(i) the ratio of HCAL to ECAL energy deposits for the leading jet, $\rhad^{j_1} \equiv E_{\rm HCAL}^{j_1} / E_{\rm ECAL}^{j_1}$;
(ii) the number of charged subparticles inside the leading jet, $N_{\rm charged}$;
(iii) the minimum angular separation between the reconstructed $W$ boson and the leading jet or photon,  $\min\{ \Delta R(W, \gamma_1), \Delta R(W, j_1) \}$;
(iv) the ratio $\pt^j / H_T$, where $H_T$ is the scalar sum of the transverse momenta of all hadronic activity in the event.

\begin{table*}[t]
  \centering
 \setlength{\tabcolsep}{5pt}
  \renewcommand{\arraystretch}{1.2}
  {\footnotesize
  \begin{tabular}{|l||c|c|c|c|c|c|c|}
  \hline
\multicolumn{8}{|c|}{Cut-flow for the signal $pp\to \ch \hf \to \ell^\pm \nu \, \rcol \, \egmj$  at the 14 TeV LHC}
\\ \hline
 & \multirow{2}{*}{Background} & \multicolumn{6}{c|}{ $\mhf=0.5\gev,~\tb=10$} \\ \cline{3-8}
 &  & \multicolumn{2}{c|}{ $\mch=100\gev$} &  \multicolumn{2}{c|}{ $\mch=200\gev$} &  \multicolumn{2}{c|}{ $\mch=300\gev$} \\ \cline{2-8}
 & $\sigma_{\rm bg}$ [fb] & $\sigma_{\rm sg}$ [fb] & $\mathcal{S}^{10\%}_{3 \iab}$
& $\sigma_{\rm sg}$ [fb] & $\mathcal{S}^{10\%}_{3\iab}$
& $\sigma_{\rm sg}$ [fb] & $\mathcal{S}^{10\%}_{3 \iab}$\\ \hline
 Basic & $1.93\times 10^2$ & $1.09 \times 10^0$ & 0.056 & $4.83 \times 10^{-1}$ & 0.025 & $1.88 \times 10^{-1}$ & 0.010 \\  \hline
 $\rhad^{j_1}>950$ & $ 4.31  \times 10^0$  & $1.47\times 10^{-1} $ & 0.336 & $2.69 \times 10^{-1}$ & 0.610 & $1.24 \times 10^{-1}$ & 0.283 \\ \hline
 $N_{\rm charged} \leq 1$ & $3.34 \times 10^{-1}$ & $1.25 \times 10^{-1}$  & 3.208 &
 $2.64 \times 10^{-1}$ & 6.206 & $1.22\times 10^{-1}$ & 3.138 \\
 \hline
 $\min \Dt R_W<2$ & $1.54 \times 10^{-1}$ & $1.18 \times 10^{-1}$ & 5.769 &
 $ 1.84 \times 10^{-1}$ & 8.341 & $ 5.08 \times 10^{-2}$ & 2.738 \\ \hline
 $\pt^{j_1}/H_T > 0.3$ & $6.15  \times 10^{-2}$ & $ 7.34  \times 10^{-2}$ & 7.555 & $1.31  \times 10^{-1}$ & 11.954
 & $3.59 \times 10^{-2}$ & 4.102 \\ \hline
  \end{tabular}
  }
  \caption{Cut-flow for the signal process $pp \to \ch \hf \to \ell^\pm \nu + \rcol + \egmj$ at the 14 TeV LHC, along with the dominant background process $pp \to \ell^\pm \nu j \gamma$. Results are shown for three benchmark scenarios with  $\mhf=0.5\gev$, $\tb = 10$, and $\mch = 100, 200, 300\gev$. The variable $\min \Delta R_W$ denotes the minimum angular separation between the reconstructed $W$ boson and the leading jet or photon: $\min \{ \Delta R(W, j), \Delta R(W, \gamma) \} $. The  significance is computed assuming a 10\% background background uncertainty and an integrated luminosity of $3\iab$. 
   }
  \label{tab:cutflow_HLLHC}
\end{table*}

The discriminating power of these four variables is illustrated in \autoref{tab:cutflow_HLLHC}, which presents the cut-flow results for the signal process $pp \to \ch \hf \to \ell^\pm \nu \, \rcol \, \egmj$ 
and the dominant background $pp\to\ell^\pm\nu j \gm $ at the 14 TeV LHC. Three benchmark scenarios are considered for the signal, corresponding to $\mhf=0.5\gev$, $\tb = 10$ and $\mch = 100, 200, 300\gev$. The table demonstrates how each successive cut impacts the signal and background cross sections, as well as the resulting significance. The significance is calculated assuming a 10\% background uncertainty on the background and an integrated luminosity of $3\iab$, based on the profile likelihood method from Ref.~\cite{Cowan:2010js}:
\bea
\label{eq-significance:nbg}
\mathcal{S} &=& 
\left[2(\nsg + \nbg) \log\left(\frac{(\nsg + \nbg)(\nbg + \dbg^2)}{\nbg^2 + (\nsg + \nbg)\dbg^2} \right)  
%\right. \\ \nn && \left.
- 
\frac{2 \nbg^2}{\delta_b^2} \log\left(1 + \frac{\dbg^2 \nsg}{\nbg (\nbg + \dbg^2)}\right)\right]^{1/2},
\eea
where $\nsg$ denotes the number of signal events, 
$\nbg$ the number of background events, and $\dbg = \Dbg  \nbg$ the background uncertainty yield.
In the following discussion, the selection efficiency of each specific cut is measured relative to the preceding cut in the cut-flow.

Let us first discuss the variable $\rhad^{j_1}$.
A typical QCD jet with $\pt > 70\gev$, as required at the basic selection level, deposits a significant fraction of its energy in the ECAL due to the abundance of neutral pions, which promptly decay into photon pairs.
In contrast, our signal events feature an emerging photon jet in the HCAL, which does not produce energy deposits in the ECAL, resulting in a large $\rhad^{j_1}$.
Imposing the requirement $\rhad^{j_1} > 950$ proves very efficient, retaining only approximately 2\% of background events while preserving approximately 13.5\%, 55.7\%, and 66.0\% of signal events for $\mch=100,200,300\gev$, respectively.\footnote{A non-negligible fraction of signal events have small $\rhad^{j_1}$, particularly pronounced for lighter $\mch$. This occurs because, for light $\mch$, the emerging photon jet originating from the charged Higgs decay ($\ch\to\wpm \hf$) can be too soft to pass the $\pt>70\gev$ threshold imposed in the basic selection, leading the ISR jet to become the leading jet candidate instead.}
Nevertheless, the significances after this initial cut are far from sufficient to claim a confident detection.

The second variable, the number of charged subparticles within the leading jet, is motivated by a key signal feature: the diphoton jet arising from $\hf$ decay contains no charged tracks. Imposing the requirement $N_{\rm charged} \leq 1$ substantially reduces the background, retaining only about 7.7\% of background events, while preserving high signal efficiencies of 85.0\%, 98.1\%, and 98.4\% for $\mch = 100, 200, 300\gev$, respectively. This variable proves highly effective, enhancing the signal significance by nearly an order of magnitude. The resulting significances reach $3.2\sg$, $6.2\sg$, and $3.1\sg$ for $\mch = 100, 200, 300\gev$, respectively.

The third variable is the minimum of two angular separations: $\Delta R(W, \gamma)$ and $\Delta R(W, j)$. The $W$ four-momentum is reconstructed from the $\ell \nu$ final state, following the standard procedure described in Refs.~\cite{ATLAS:2015pfy,CMS:2018quc,ATLAS:2018fwq,ATLAS:2019hxz,CMS:2021vhb}. In the signal, the $W$ boson arises from the decay of the charged Higgs boson and is accompanied by an $\hf$ that subsequently decays into either a $\rcol$ or an $\egmj$. Consequently, the $W$ boson is expected to be spatially close to either the photon or the jet (see \autoref{fig-Feynman}). Imposing the condition $\min \{ \Delta R(W, j), \Delta R(W, \gamma) \} < 2$ yields a significant improvement in sensitivity. In particular, for $\mch = 100$ and $200\gev$, the resulting significances rise well above the discovery threshold, reaching $5.8\sg$ and $8.3\sg$, respectively.\footnote{For $\mch = 300\gev$, the condition $\min \{ \Delta R(W, j), \Delta R(W, \gamma) \} < 2$ slightly reduces the significance by about 13\%. This occurs because heavier charged Higgs bosons are less boosted compared to lighter ones, causing their decay products to be more widely separated. }

The fourth variable, $\pt^j/H_T$, provides additional discrimination between signal and background. In our signal events, which contain no colored final-state particles, the total hadronic activity ($H_T$) tends to be relatively low. In contrast, background events involving genuine QCD jets typically exhibit higher $H_T$ due to their substantial hadronic content. Imposing the condition $\pt^j/H_T > 0.3$ modestly enhances the sensitivity, yielding final significances of $7.6\sg$, $12.0\sg$, and $4.1\sg$ for $\mch = 100, 200, 300\gev$, respectively. The first two benchmarks clearly exceed the discovery threshold. Although the $\mch = 300\gev$ case falls slightly short of $5\sg$, the resulting $4.1\sg$ significance still constitutes strong evidence for the signal. The reduced sensitivity in this case is primarily due to the smaller production cross section associated with the heavier charged Higgs boson.

\begin{figure}[!t]
\centering
\includegraphics[width=0.6\textwidth]{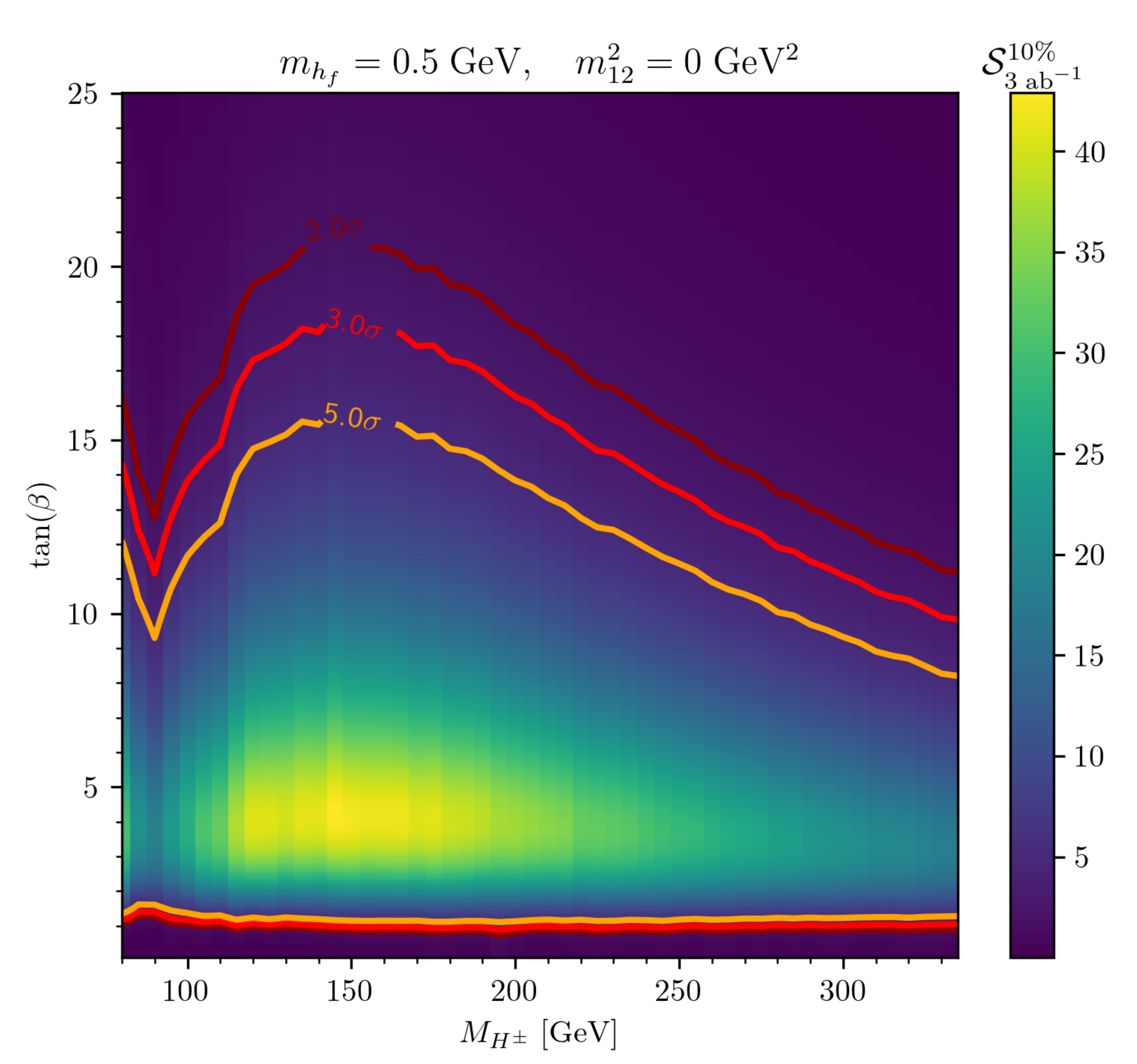} 
\vspace{-0.2cm}
\caption{Signal significances (indicated by color shading) in the allowed $(\mach, \tb)$ parameter space, with $\mhf = 0.5\gev$. Contours corresponding to $5\sigma$, $3\sigma$, and $2\sigma$ significances are shown as orange, red, and brown lines, respectively.}
\label{fig-final-significance} 
\end{figure}

Building on our dedicated strategy to probe the ultralight $\hf$ through the emerging photon jet in the HCAL, we compute the signal significance across the entire allowed parameter space. The results for $\mhf = 0.5\gev$ are shown in \autoref{fig-final-significance}, with significance values encoded by the color scale. The $5\sigma$, $3\sigma$, and $2\sigma$ contours are indicated by orange, red, and brown lines, respectively.

The results are highly encouraging. In the region $\tb \in [3, 5]$ and $\mch \in [120, 190]\gev$, the signal significance exceeds $40\sigma$.
In particular, the entire allowed range of $\mch$ supports $5\sigma$ discovery potential.
The minimum $\tb$ along the $5\sigma$ contour is about 1.5, while the maximum $\tb$ rises with $\mch$ up to $\sim 15$ at $\mch \sim 150\gev$ and then decreases to $\sim 8$ at $\mch \sim 335\gev$.
In summary, even under conservative assumptions for background uncertainties, the discovery of the ultralight $\hf$ via the emerging photon jet signature in the HCAL is well within reach at the HL-LHC.

\section{Conclusion} \label{sec-conclusion}

We have proposed a novel collider signature for neutral long-lived particles: the emerging photon jet in the hadronic calorimeter (HCAL). This distinctive signal arises when a collimated photon pair, produced from the decay of an ultralight scalar, initiates an electromagnetic shower deep in the HCAL, without associated charged tracks or ECAL deposits. To our knowledge, this is the first dedicated study of such a signature.

As a concrete realization, we examined the ultralight fermiophobic Higgs boson $\hf$ in the Type-I two-Higgs-doublet model. In the golden channel $pp \to \ch \hf \to W^\pm \hf \hf$, we focused on events featuring one $\hf$ decaying in the ECAL (as a single photon) and the other in the HCAL (as an emerging photon jet). Our detector-level analysis shows that this signature provides discovery-level sensitivity across a wide region of theoretically and experimentally allowed parameter space.

While demonstrated in the context of the fermiophobic Type-I 2HDM, the emerging photon jet in the HCAL offers a broadly applicable and previously unexplored strategy for probing neutral LLPs decaying into photons—highlighting a promising new direction in long-lived particle searches at collider experiments.

%\bibliographystyle{JHEPMod}

%\bibliography{emerging-photon}  

\end{document}